\documentclass[aps,prd,twocolumn,nofootinbib]{revtex4-2}
\usepackage{amsmath,amssymb,bm,hyperref}
\usepackage[utf8]{inputenc}
\usepackage[T1]{fontenc}
\usepackage{graphicx}
\usepackage{color,caption}

\begin{document}
	
	\title{Entropy Quantization and Quasi-normal Modes of Dyonic Kerr-Sen Black Holes}
	
	\author{Muhammad Fitrah Alfian Rangga Sakti$^{1,2,3}$}
	\email{fitrahalfian@gmail.com}
	\author{Piyabut Burikham$^{1}$}%
	\email{piyabut@gmail.com}
	\affiliation{$ ^1 $High Energy Physics Theory Group, Department of Physics,
		Faculty of Science, Chulalongkorn University, Bangkok 10330, Thailand,}
	\affiliation{$ ^2 $Department of Physics and Astronomy, University of Waterloo, Waterloo, Ontario, N2L 3G1, Canada,}
	\affiliation{$ ^3 $Perimeter Institute for Theoretical Physics, Waterloo, Ontario, N2L 2Y5, Canada.}
	
	\begin{abstract}
		We explore the properties of inner and outer horizon thermodynamics of dyonic Kerr-Sen black hole (DKSBH). It is observed that the entropy (or area) product is universal, depending only on the angular momentum of the BH. We then proceed to study the dual conformal field theory~(CFT) in the Kerr/CFT correspondence using thermodynamic relations and compute the central charges from 2D CFT. The central charges are found to be universal with only angular momentum dependence. By comparing to Kerr-Newman BH, it is found that the essential difference is in the right-moving sector of the CFT. Interestingly, we can then explicitly produce the non-vanishing central charges related to its static solution, the dyonic dilaton BH, using the thermodynamic method. Moreover, from the CFT relations to multi-horizon thermodynamics, we find the analytical expression of the quasi-normal modes~(QNM) spectra in terms of BH parameters. This QNM computation supports the Bekenstein-Hod conjecture on the BH's entropy quantization.
	\end{abstract}
	
	\maketitle
	
	\section{Introduction}
	
	When the classical general relativity theory was found decades ago, Hawking and Bekenstein separately conjectured that any BH in thermal equilibrium has an entropy and a temperature. The entropy is known to be solely proportional to the quarter area of the BH. The exact expression of the entropy of the BH on the event horizon (at $r=r_+$) is explicitly given by \cite{Hawking1975,Bekenstein1972,Bekenstein1973,Bekenstein1974}
	\begin{equation}
		S_+ = \frac{k _BA_+}{4l^2_{P}},
	\end{equation}
	where the Planck length $l^2_P = G\hbar/c^3$ and $A_+$ is the area of the event horizon. The BH temperature is proportional to the surface gravity at $r_+$ as given by
	\begin{equation}
		T_+ = \frac{\hbar\kappa_+}{2\pi k_B}.
	\end{equation}
	where $\kappa_+$ denotes the surface gravity of the BH on $r_+$. 
	
	The presence of $l^2_{P}$ is the result coming from the non-perturbative quantum gravity. Moreover, the presence of Boltzmann constant $k_B$ shows the point of view of the entropy as the logarithm of density of states of thermodynamical system. The computation done by Hawking and Bekenstein provides us the relation between quantum and gravitational theories represented by BHs.

	Since the above formulation of entropy and temperature occurs on the event (outer) horizon $r_+$, it is natural to question whether such relations are also valid on the Cauchy~(inner) horizon (at $r_-$). Now it is acclaimed that such relations also hold on $r_-$, resulting the following similar formula,
	\begin{equation}
		S_- = \frac{k _BA_-}{4l^2_{P}}, ~~~T_- = \frac{\hbar\kappa_-}{2\pi k_B},
	\end{equation}
	where $A_-$ and $\kappa_-$ are the area and surface gravity on the inner horizon.
	
	Some speculations about the the area of the BHs is being quantized appears in Refs. \cite{BekensteinLNC1974,MukhanovJETPL1986,BekensteinMukhanovPLB1995,HodPRL1998} while it is also supported by the calculation for some specific BH solutions, for example, in string theory and M-theory \cite{CveticYoumPRD1996,CveticLarsenNRB1997,CveticLarsenPRD1997}. For asymptotically flat BPS BHs in four and higher dimensions, the quantization rule becomes
	\begin{equation}
		S_\pm = 2\pi \ell_{\rm pl}^2\left(\sqrt{N_1}\pm \sqrt{N_2}\right),
	\end{equation}
	or one can find the entropy product as \cite{LarsenPRD1997,VisserPRD2013,CastroRodriguezPRD2012,DetournayPRL2012,FaraoniMorenoPRD2013,ChenLiuJHEP2012,ChenXueJHEP2013,ChenZhangJHEP2013,CastroDehmamiJHEP2013,PradhanEPJC2014}
	\begin{equation}
		S_+S_- = (2\pi \ell_{\rm pl}^2)^2 N, ~~~N, N_1, N_2\in \mathbb{N}.\label{eq:entropyproduct}
	\end{equation}
	As we can observe that the above relations are mass independent or universal. Such universal thermodynamic products have attracted considerable attention because they provide direct probes of the microscopic structure of BH entropy via BH/CFT correspondence.
	
	The Kerr-Sen BH \cite{SenPRL1992} is an exact rotating charged solution in four-dimensional low-energy heterotic string theory which carries mass $M$, angular momentum $J$, and electric charge $Q$. It reduces to the Kerr BH in the uncharged limit. In the Einstein frame, the entropy (area) product is universal while the entropy (area) sum depends on the ADM mass \cite{PradhanEPJC2016}.
	
	The dyonic generalization of the Kerr-Sen BH arises naturally in
	Einstein-Maxwell-Dilaton-Axion (EMDA) theory, where the BH simultaneously
	carries both electric charge $Q$ and magnetic charge $P$ \cite{WuWuWuYuPRD2021}. This
	dyonic Kerr-Sen BH has attracted growing interest in recent
	years \cite{SaktiPiyabutPRD2022,SaktiEPJC2023,JanaKarPRD2023,SaktiPiyabutEPJC2024,SenjayaEPJC2024,SenjayaBurikhamEPJC2025}. Its richer charge structure introduces qualitatively new
	features at the level of multi-horizon thermodynamics: the interplay between
	$Q$ and $P$ modifies the thermodynamic properties of the solution such as Smarr and Christodoulou-Ruffini irreducible mass \cite{WuWuWuYuPRD2021}. 
	
	Recently, the investigation of holographic pictures of BHs has grown rapidly in order to find the alternative and non-perturbative way of grasping its quantum properties. It is conjectured in Ref.~\cite{GuicaPRD2009} that the BH partition function is dual to the 2D CFT partition function which leads to the matching between Bekenstein-Hawking entropy of Kerr BH and Cardy entropy formula from 2D CFT. Then there are subsequent development to study more BH solutions in Einstein \cite{SaktiSurosoIJMPD2018,SaktiSurosoEPJPlus2019} or beyond Einstein theory \cite{SaktiPhysDarkU2021}. Furthermore, the radial wave equation on BH's background also captures hidden conformal symmetry \cite{CastroPRD2010} that generalizes the result from \cite{GuicaPRD2009} for both sectors in 2D CFT. For some development of the hidden conformal symmetry, see Refs.~\cite{SaktiGREG2019,SaktiNucPhysB2020,SaktiPhysDarkU2022}.
	
	Another interesting perspective of the dual CFT of BH is to consider the 2D CFT thermodynamics from a BH thermodynamic in multi-horizon perspective \cite{ChenLiuJHEP2012,ChenXueJHEP2013,ChenZhangJHEP2013,GolchinJHEP2020,YektaGolchinEPJC2020}. It has opened a new window and shed considerable light on the black hole/CFT correspondence. It has long been recognized that the inner horizon may play a pivotal role in the microscopic counting of entropy associated with the outer horizon. In particular, it was pointed out in Refs.~\cite{ChenLiuJHEP2012,ChenXueJHEP2013,ChenZhangJHEP2013} that the mass-independence of the entropy product $\mathcal{S}_{+}\mathcal{S}_{-}$ carries non-trivial implications for the holographic description of black holes. The mass-independence of the entropy product $\mathcal{S}_{+}\mathcal{S}_{-}$ is precisely equivalent to the condition
	\begin{equation}
		T_{+}\mathcal{S}_{+} + T_{-}\mathcal{S}_{-}=0.\label{eq:TSuniversality}\footnote{
			Note that in some papers such as \cite{ChenLiuJHEP2012,ChenXueJHEP2013,ChenZhangJHEP2013}, they use $T_{+}\mathcal{S}_{+} =T_{-}\mathcal{S}_{-}$ since they cast out the negative sign on $T_-$.		}
	\end{equation}
	Interestingly, building upon these foundational results, it was proposed also that this thermodynamic method can be applied successfully to certain regular BHs \cite{GolchinPRD2022} and acoustic BHs \cite{AnacletoPLA2016}.
	
	In the present work, we explicitly extend this multi-horizon thermodynamic framework to the dyonic Kerr-Sen black hole, which carries both electric charge $Q$ and magnetic charge $P$ in addition to mass $M$ and angular momentum $J$. We demonstrate that the entropy product $\mathcal{S}_{+}\mathcal{S}_{-}$ for the DKSBH is mass-independent, thereby satisfying the universality criterion. We use this property to derive the central charges $c_{L} = c_{R}$ of the dual two-dimensional CFT in the Kerr/CFT correspondence via the Cardy entropy formula. These results provide further microscopic evidence for the holographic description of the entropy of the DKSBH. We briefly compare the results from thermodynamic method of the dyonic Kerr-Sen BH to those of Kerr-Newman BH. Moreover, we also derive the CFT temperatures and central charge of dyonic dilaton BH which is the static version of dyonic Kerr-Sen BH.
	
	By studying the radial scalar wave on dyonic Kerr-Sen background with low-frequency limit, we find exactly the same results of the CFT temperature derived from thermodynamic method. Furthermore, the existence of dual CFT for dyonic Kerr-Sen BH allows us to compute the QNM spectra in the dyonic Kerr-Sen background due to the fact that the radial scalar wave equation captures $SL(2,R)\times SL(2,R)$ isometry.
	
	The paper is organized as follows. Section~\ref{sec:BHsolution} presents the dyonic Kerr-Sen BH solution and its thermodynamics. Section~\ref{sec:entropyquant}
	derives the thermodynamic products and exhibits the universal properties of dyonic Kerr-Sen BH thermodynamics. In Section~\ref{sec:CFT}, we compute the CFT thermodynamic quantities and subsequently compare it with those of Kerr-Newman BH. We also compute the CFT thermodynamics of dyonic dilaton BH in Section \ref{sec:static} and the QNM specturm from the dual CFT in Section \ref{sec:QNM}. Then in Section \ref{sec:BekensteinHod}, we explain how the QNM spectra correspond to the entropy quantization based on Bekenstein-Hod conjecture. Finally, Section \ref{sec:summary} summarizes our results. Throughout this work, the natural units $G = \hbar = c = k_B = 1$ are used.

	\section{Dyonic Kerr-Sen Black Hole}
	\label{sec:BHsolution}
	\subsection{Spacetime metric}
	\label{subsec:spacetime}
	An exact rotating dyonically and electrically charged BH solution in four dimensional Einstein-Maxwell-Dilaton-Axion theory is given by the following metric in Boyer-Lindquist coordinates \cite{WuWuWuYuPRD2021},
	\begin{equation}
		ds^2 = - \frac{\Delta}{\varrho^2 } X^2+  \frac{\varrho ^2}{\Delta}dr^2 + \varrho ^2 d\theta^2 + \frac{\sin^2\theta}{\varrho ^2}Y^2,\label{eq:dKSmetric} \
	\end{equation}
	where
	\begin{equation}
		X = dt - a \sin^2\theta d\phi, ~~ Y = adt- (r^2-2dr-k^2+a^2 )d\phi, 
	\end{equation}
	\begin{equation}
		\Delta = r^2-2dr-2m(r-d) -k^2+a^2+p^2+q^2, 
	\end{equation}
	\begin{equation}
		\varrho^2 = r^2-2dr -k^2 + a^2\cos^2\theta.\label{eq:metricfunctiondKS}\
	\end{equation}
	The parameters $m, a, q, p, d, k$ are mass, spin, electric charge, magnetic (dyonic) charge, dilaton charge, and axion charge of the black hole, respectively. The dilaton and axion charges explicitly depend on the electromagnetic charges with the following relations
	\begin{equation}
		d= \frac{p^2 -q^2}{2m}, ~~~~~ k =\frac{pq}{m}.\
	\end{equation}
	As we can observe, when one of the electromagnetic charges vanishes, the axion charge will also vanish. When both electromagnetic charges possess the same value, the dilaton charge will vanish.
	
	The electromagnetic field, its dual, dilaton scalar, and axion pseudoscalar fields of the dyonic Kerr-Sen BH (\ref{eq:dKSmetric}) are given in the following exquisite forms
	\begin{equation}
		\textbf{A} = \frac{q(r-p^2/m)}{\varrho^2}X-\frac{p\cos\theta}{\varrho^2}Y,\label{eq:AdKS}\ \end{equation}
	\begin{equation}
		\textbf{B} = \frac{p(r-p^2/m)}{\varrho^2}X+\frac{q\cos\theta}{\varrho^2}Y, \label{eq:BdKS}\
	\end{equation}
	\begin{equation}
		e^{\phi} = \frac{r^2+(k+a\cos\theta)^2}{\varrho^2}, \label{eq:dilatondKS}
	\end{equation}
	\begin{equation}
		\chi = 2\frac{kr-d(k+a\cos\theta)}{r^2+(k+a\cos\theta)^2}. \label{eq:axiondKS}
	\end{equation}
    It is worth to emphasize that in Ref. \cite{WuWuWuYuPRD2021}, a concrete gauge choice has been chosen to ensure that the temporal components of the electromagnetic potentials vanishes at infinity. Moreover, to make the expressions more symmetric, one can work with another radial coordinate by shifting $r\rightarrow r+p^2/m$ or $r\rightarrow r+d$ \cite{WuWuWuYuPRD2021}. However, we will not shift the radial coordinate throughout this work. For non-dyonic solution, one can use results in \cite{WuWuYuPRD2020} and turn off $p$. The original derivation for Kerr-Sen BH is given in \cite{SenPRL1992}.
	
\begin{figure}
		\centering
		\includegraphics[width=0.9\columnwidth]{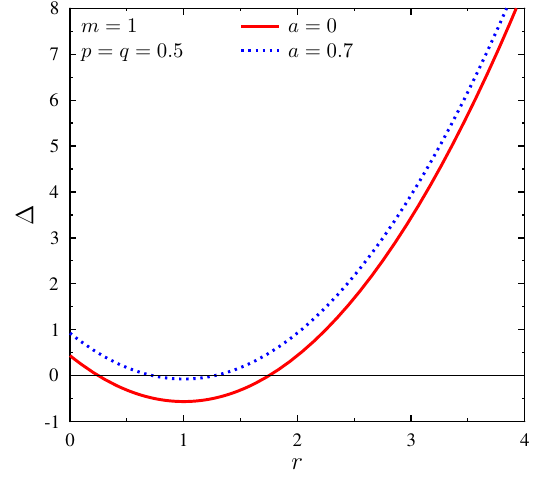}
		\caption{Plot of $\Delta$ as a function of $r$ with $m=1, p=q=0.5$ and variation $a=0, 0.7$.}
		\label{fig:Horizon}
	\end{figure}

	There are two horizons for dyonic Kerr-Sen BH, namely event horizon ($r_+$) and Cauchy horizon ($r_-$). Their radii can be determined by solving $\Delta=0$ as the horizons are given by
	\begin{equation}
		r_\pm = \left(m+d\right)\pm \sqrt{m^2+d^2+k^2-p^2-q^2-a^2}.\label{eq:rpm}
	\end{equation}
	The plot of $\Delta$ as a function of $r$ is given in Fig. \ref{fig:Horizon} which shows that this black hole possesses two horizons, even for static case ($a=0$).
	From Eq. (\ref{eq:rpm}), the requirement for the BH to have horizons is when
	\begin{equation}
		a^2+p^2+q^2 \leq m^2+d^2+k^2.\label{eq:nonakedsing}
	\end{equation}
	Hence, the extremal limit of the dyonic Kerr-Sen BH corresponds to
	\begin{equation}
		a^2+p^2+q^2 = m^2+d^2+k^2, ~~~ r_\pm=m+d.\label{eq:extremalcon}\
	\end{equation}
	
	\subsection{Thermodynamics}
	\label{subsec:thermo}
	The thermodynamic quantities and properties on $r_+$ of this BH are derived in \cite{WuWuWuYuPRD2021}. For the thermodynamic quantities and properties on $r_-$, one can replace $r_+ \rightarrow r_-$. Thus, we can obtain the thermodynamic quantities on both horizons. The physical mass, electric charge, magnetic charge, and angular momentum are given by
	\begin{equation}
		M=m,~ Q= q, ~ P = p, ~J=ma.\label{eq:MQPJ}
	\end{equation}
	The area of this BH is explicitly
	\begin{equation}
		A_\pm = 4\pi(r_\pm^2 - 2dr_\pm -k^2 +a^2).
	\end{equation}
	Thus the entropy is given by
	\begin{equation}
		S_\pm = \pi(r_\pm^2 - 2dr_\pm -k^2 +a^2).\label{eq:entropy}
	\end{equation}
	
	The angular velocity on the horizon is
	\begin{equation}
		\Omega_\pm = \frac{a}{r_\pm^2 - 2dr_\pm -k^2 +a^2},\label{eq:angvelocity}
	\end{equation}
	and the surface gravity is given by
	\begin{equation}
		\kappa_\pm = \frac{r_\pm-(m+d)}{r_\pm^2 - 2dr_\pm -k^2 +a^2}.\label{eq:surface}
	\end{equation}
	The temperature can be expressed in terms of surface gravity which reads as
	\begin{equation}
		T_\pm = \frac{\kappa_\pm}{2\pi}.\label{eq:temperature}
	\end{equation}
	It is worth noting that $T_+>T_-$ and $T_-<0$ 	\footnote{Note that $T_-<0$ is our convention which is different from Refs. \cite{ChenLiuJHEP2012,ChenXueJHEP2013,ChenZhangJHEP2013} where they cast out the negative sign. Thus in their work, $T_- \rightarrow -|T_-|$ or on the other hand, $\kappa_- \rightarrow -|\kappa_-|$. If one wants to cast out the negative sign, one must consistently do it in all equations containing $T_-$ or $\kappa_-$. Explicitly, in addition to the universal relation (\ref{eq:TSuniversality}) that have changed as given in footnote 1, casting out the negative sign will change Eq. (\ref{eq:OmegaTuniversality}) to $\frac{\Omega_+}{T_+}=\frac{\Omega_-}{T_-}$, the terms $T_- S_-, \frac{\kappa_-}{8\pi} dA_-, 2 T_- S_-$ in Eqs. (\ref{eq:firstlaw})-(\ref{eq:Smarr}) to  $-T_- S_-, -\frac{\kappa_-}{8\pi} dA_-, -2 T_- S_-$, respectively. It will also change the term $\frac{\kappa_-}{8\pi}$ in lhs of Eq. (\ref{eq:kappaterm}) to $-\frac{\kappa_-}{8\pi}$, lhs of Eq. (\ref{eq:TT1}) to $-T_-T_+$, and lhs of Eq. (\ref{eq:TT2}) to $T_+ - T_-$.}
	
		The first law of thermodynamics of dyonic Kerr-Sen BH is 
	\begin{equation}
		dM = T_\pm S_\pm + \Omega_\pm dJ + \Phi_\pm dQ +\Psi_\pm dP.\label{eq:firstlaw}
	\end{equation}
	The first law can also be rewritten in terms of black hole's area, the surface gravity and the potentials,
	\begin{equation}
		dM = \frac{\kappa_\pm}{8\pi} dA_\pm + \Omega_\pm dJ + \Phi_\pm dQ +\Psi_\pm dP. \label{eq:firstlaw1}
	\end{equation}
	It is easy to derive the Smarr formula that is given by
	\begin{equation}
		M =2 T_\pm S_\pm +2\Omega_\pm J + \Phi_\pm Q +\Psi_\pm P,\label{eq:Smarr}
	\end{equation}
	and the Christodoulou-Ruffini square mass formula explicitly reads as
	\begin{equation}
		M^2 = \frac{A_\pm}{16\pi}+\frac{4\pi J^2}{A_\pm}+\frac{P^2+Q^2}{2}.\label{eq:Christodouloumass}
	\end{equation}
	One can derive the following expressions for the thermodynamic quantities,
	\begin{equation}
		\frac{\kappa_\pm}{8\pi} = \frac{\partial M}{\partial A_\pm} =\frac{1}{M}\left(\frac{1}{32\pi} -\frac{2\pi J^2}{A_\pm^2}\right),\label{eq:kappaterm}
	\end{equation}
	\begin{equation}
		\Omega_\pm = \frac{\partial M}{\partial J} = \frac{4\pi J}{M A_\pm},\label{eq:omegaterm}
	\end{equation}
	\begin{equation}
		\Phi_\pm = \frac{\partial M}{\partial Q} = \frac{Q}{2M},\label{eq:ElPot}
	\end{equation}
	\begin{equation}
		\Psi_\pm = \frac{\partial M}{\partial P} = \frac{P}{2M}.\label{eq:MagPot}
	\end{equation}
	Notably, the electric and magnetic potentials are the same for both inner and outer horizons.
	
	\section{Entropy Quantization}	
	\label{sec:entropyquant}
	
	It has been observed that the product of horizon entropies for many BH solutions is mass independent as given in Eq. (\ref{eq:entropyproduct}) where $N$ depends only on the quantized charges like angular momentum	and electric charge. This universality encodes the quantum nature of the entropy. Motivated by this universality, we will derive the similar relation for dyonic Kerr-Sen BH which will benefit us to study the dual CFT in the next section.
	
	Firstly, we start by computing the product and sum of the entropies which are given by
	\begin{equation}
		S_-S_+ = (2\pi J)^2. \label{eq:entropyproductdKS}
	\end{equation}
	and
	\begin{equation}
		S_-+S_+ = 2\pi(2M^2-Q^2-P^2). \label{eq:entropysumdKS}
	\end{equation}
	As we can see that the entropy product is quantized which depends on the angular momentum only. It is somehow different from the dyonic Kerr-Newman BH where the entropy product also depends on the charges. The universal entropy product implies that we can derive the CFT thermodynamics from this relation.	One may also compute from the square mass formula (\ref{eq:Christodouloumass}) that the universal area product satisfies
		\begin{equation}
			A_- A_+ =(8\pi J)^2.\label{eq:}
		\end{equation}
	which also confirms the universal entropy product (\ref{eq:entropyproductdKS}).
	
%
%
%
%
%
%

The other important relation is Eq.~(\ref{eq:TSuniversality}). The relation can be straightforwardly derived from entropy (\ref{eq:entropy}) and temperature (\ref{eq:temperature}). Moreover, one can also show that
\begin{equation}
	\frac{\Omega_+}{T_+}+\frac{\Omega_-}{T_-}=0,\label{eq:OmegaTuniversality}
\end{equation}
which is also universal.


Now we want to derive the entropy bound of for both horizons which is exactly Penrose-like inequality for event horizon. From Eq. (\ref{eq:nonakedsing}), we obtain
\begin{equation}
	M^2 \geq J+\frac{P^2+Q^2}{2}.
\end{equation}
Since $r_+\geq r_-$, therefore $S_+\geq S_-\geq 0$ and from the entropy product (\ref{eq:entropyproduct}), we obtain
\begin{equation}
	S_+ \geq \sqrt{S_+S_-}=2\pi J \geq S_-.
\end{equation}
We know that 
\begin{equation}
	S_+ + S_- \geq S_+ \geq \frac{S_++S_-}{2}\geq S_-,
\end{equation}
thus we have
\begin{equation}
	2\pi(2M^2-P^2-Q^2)\geq S_+ \geq \pi(2M^2-P^2-Q^2)\geq S_-.
\end{equation}
Then we can find the entropy bound as follows
\begin{equation}
	\pi(2M^2-P^2-Q^2) \leq S_+ \leq 2\pi(2M^2-P^2-Q^2).\label{eq:entropybound+}
\end{equation}
and
\begin{equation}
	0 \leq S_- \leq 2\pi J.\label{eq:entropybound-}
\end{equation}
Note that in the limit $Q =P= 0$, we obtain the Kerr entropy bound.

The other fruitful relations beside the entropy product and sum are the temperature product and sum. It is straightforward that we can obtain
\begin{equation}
	T_-T_+ = -\frac{(2M^2-P^2 -Q^2)^2-4J^2}{(8\pi JM)^2},\label{eq:TT1}
\end{equation}
and
\begin{equation}
	T_-+T_+ = -\frac{(2M^2-P^2-Q^2)^2-4J^2}{8\pi MJ^2}.\label{eq:TT2}
\end{equation}
Both temperature product and sum are not universal.

\section{CFT Thermodynamics}
\label{sec:CFT}
In this section, we will derive the CFT thermodynamics from BH thermodynamics on $r_\pm$ from the fact that dyonic Kerr-Sen BH satisfies the universality of the entropy product. Thus we want to prove that through thermodynamic method, we can construct the dual CFT where we can derive exactly the same central charges of the left- and right-moving sectors.

In 2D CFT, the first law and energy formula for both left- and right-moving sectors are given by \cite{ChenLiuJHEP2012,ChenXueJHEP2013}
\begin{equation}
	dE_{L,R} = T_{L,R} dS_{L,R} + \Omega_{L,R} dJ + \Phi_{L,R} dQ+\Psi_{L,R} dP,
\end{equation}
\begin{equation}
	E_{L,R}= 2T_{L,R} S_{L,R} + 2\Omega_{L,R} J + \Phi_{L,R} Q+\Psi_{L,R} P,
\end{equation}
respectively. The relations between the BH thermodynamics and 2D CFT thermodynamics read as\footnote{If one wants to cast out the negative sign of $T_-$, it will change how BH thermodynamics relates  to CFT thermodynamics such as $\frac{\Omega_+}{T_+}\pm\frac{\Omega_-}{T_-}$ to $\frac{\Omega_+}{T_+}\mp\frac{\Omega_-}{T_-}$ in Eq. (\ref{eq:massOmrel}), $\frac{1}{T_+}\pm \frac{1}{T_-}$ to $\frac{1}{T_+}\mp \frac{1}{T_-}$ in Eq. (\ref{eq:TSrel}), $\frac{\Phi_+}{T_+}\pm\frac{\Phi_-}{T_-}$ to $\frac{\Phi_+}{T_+}\mp\frac{\Phi_-}{T_-}$, $\frac{\Psi_+}{T_+}\pm\frac{\Psi_-}{T_-}$ to $\frac{\Psi_+}{T_+}\mp\frac{\Psi_-}{T_-}$ in Eq. (\ref{eq:PhiPsirel}), and lhs of Eq. (\ref{eq:scalefactor}) to $\frac{T_+^2-T_-^2}{T_- T_+ (\Omega_+ -\Omega_-)}$.} \cite{ChenLiuJHEP2012,ChenXueJHEP2013}
\begin{equation}
	E_{L,R}= \frac{M}{2}, ~~~\frac{2\Omega_{L,R}}{T_{L,R}}=\frac{\Omega_+}{T_+}\pm\frac{\Omega_-}{T_-},\label{eq:massOmrel}
\end{equation}
\begin{equation}
	\frac{1}{T_{L,R}}= \frac{1}{T_+}\pm \frac{1}{T_-}, ~~~ S_{L,R}=\frac{S_+ \pm S_-}{2},\label{eq:TSrel}
\end{equation}
\begin{equation}
	\frac{2\Phi_{L,R}}{T_{L,R}}=\frac{\Phi_+}{T_+}\pm\frac{\Phi_-}{T_-},~~\frac{2\Psi_{L,R}}{T_{L,R}}=\frac{\Psi_+}{T_+}\pm\frac{\Psi_-}{T_-}.\label{eq:PhiPsirel}
\end{equation}

Using the above relations, we can find
\begin{equation}
	T_L = \frac{1}{8\pi M},~~	T_R = \frac{\sqrt{\left(M^2-\frac{(P^2+Q^2)}{2}\right)^2-J^2}}{4\pi M(2M^2-P^2-Q^2)}\label{eq:TLR}
\end{equation}
\begin{equation}
	S_L = \pi(2M^2-P^2-Q^2),\label{eq:SL}
\end{equation}
\begin{equation}
	S_R = 2\pi \sqrt{\left(M^2-\frac{(P^2+Q^2)}{2}\right)^2-J^2},\label{eq:SR}
\end{equation}
\begin{equation}
	\Omega_L = 0, ~~~	\Omega_R = \frac{J}{2M(2M^2-P^2-Q^2)}, \label{eq:OmegaLR}
\end{equation}
\begin{equation}
	\Phi_{L,R}= \frac{Q}{4M}, ~~ 	\Psi_{L,R}= \frac{P}{4M}.\label{eq:PhiPsiLR}
\end{equation}

The left and right temperatures from (\ref{eq:TLR}) are of dimension $\mathcal{L}^{-1}$ and proportional to the microscopic temperatures obtained from the hidden conformal symmetry in the low-frequency scattering, up to a scale factor. The scale factor can be thought of as the size of the box in which the microscopic CFT lives. It can be determined  from the following relation \cite{ChenLiuJHEP2012,ChenXueJHEP2013,ChenZhangJHEP2013}
\begin{equation}
	R=\frac{T_-^2 -T_+^2}{T_- T_+ (\Omega_+ -\Omega_-)}.\label{eq:scalefactor}
\end{equation}
Thus for dyonic Kerr-Sen BH, we have
\begin{equation}
	R= \frac{2M(2M^2-P^2-Q^2)}{J}.\label{eq:scalefactor1}
\end{equation}
Applying the scale factor to temperatures (\ref{eq:TLR}) as $\bar{T}_{L,R}=R T_{L,R}$, the dimensionless left- and right-moving CFT temperatures are then given by
\begin{equation}
	\bar{T}_L = \frac{2M^2-P^2-Q^2}{4\pi J}, ~~~\bar{T}_R= \frac{\sqrt{\left(M^2-\frac{(P^2+Q^2)}{2}\right)^2-J^2}}{2\pi J}. \label{eq:dimTLR}\
\end{equation}

The central charges in left- and right-moving sectors of the 2D CFT can be computed via the Cardy formula as
\begin{equation}
	S_{L,R} = \frac{\pi^2}{3}\bar{c}_{L,R}\bar{T}_{L,R}.
\end{equation}
Therefore the dimensionless central charges of dual CFT are given as follows
\begin{equation}
	\bar{c}_{L,R} = 12J. \label{eq:cLcR}
\end{equation}
which is exactly the same as Kerr-Sen, Kerr-Newman, and Kerr BHs. This results show that dyonic Kerr-Sen BH is dual to a $\bar{c}_{L,R}=12J$ of 2D CFT at temperature $(T_L,T_R)$. If one applies directly the left- and right-moving temperatures (\ref{eq:TLR}) to the left- and right-entropies (\ref{eq:SL}) and (\ref{eq:SR}) without taking into account the scale factor on the temperatures, the central charges obtained are given by 
\begin{equation}
c_{L,R} = 24 M(2M^2-P^2-Q^2)\label{eq:cLcRwithdimension}
\end{equation}
which does not depend on the angular momentum explicitly.

In the extremal limit where both horizons coincide $r_+=r_-$ which is equivalent to $M^2 -\frac{P^2+Q^2}{2}= J$, the 2D CFT thermodynamic quantities reduce to
\begin{equation}
	\bar{T}_L =\frac{1}{2\pi},\qquad \bar{T}_R=0,
\end{equation}
\begin{equation}
	S_L = 2\pi J,\qquad S_R=0,
\end{equation}
\begin{equation}
	\Omega_L=0,\qquad \Omega_R = \frac{1}{4M},
\end{equation}
while $\Phi_{L,R}, \Psi_{L,R}, T_L$ have similar forms as the non-extremal condition. Therefore, we can obtain the microscopic entropy via Cardy formula in the dual CFT
\begin{equation}
	S_{CFT} = \frac{\pi^2}{3}\bar{c}_L \bar{T}_L = 2\pi J.
\end{equation}
which is in perfect agreement with macroscopic Bekenstein-Hawking entropy of the extremal dyonic Kerr-Sen BH \cite{SaktiPiyabutPRD2022}.	

\subsection{Comparison to Kerr-Newman BH}

Let us compare our dyonic Kerr-Sen BH results with Kerr-Newman BH thermodynamics given in Ref.~\cite{ChenLiuJHEP2012}. The entropy products of both dyonic Kerr-Sen BH and Kerr-Newman BH are independent of mass. However, the Kerr-Newman entropy product 
\begin{equation}
	S_{+}S_{-} = \pi^{2}(4J^{2}+Q^{4}),
\end{equation}
also depends on the charge $Q$ while the entropy product of dyonic Kerr-Sen BH (\ref{eq:entropyproductdKS}) is independent of both charges $P, Q$ and depends only on the angular momentum $J$. There is a fundamental difference between the dyonic Kerr-Sen BH and Kerr-Newman BH, turning off $P$ charge will not reduce dyonic Kerr-Sen BH to the Kerr-Newman spacetime. 

The sums of entropy, on the other hand, have very similar formula. For Kerr-Newman BH, $S_{+}+S_{-} = 2\pi (2M^{2}-Q^{2})$, while the dyonic Kerr-Sen BH entropy sum (\ref{eq:entropysumdKS}) only has extra contribution from the magnetic charge $P$ with the same dependence as the electric charge $Q$. 

For CFT thermodynamics of Kerr-Newman BH, by using (\ref{eq:TSrel}), we obtain 
\begin{eqnarray}
&&S_{L}=\pi(2M^{2}-Q^{2}),\quad S_{R}=2\pi M\sqrt{M^{2}-\left(\frac{J^{2}}{M^{2}}+Q^{2}\right)},\notag\\
&&T_{L}=\frac{1}{8\pi M},\quad T_{R}=\frac{\sqrt{M^2 -\left(\frac{J^{2}}{M^{2}}+Q^{2}\right)}}{4\pi(2 M^2 - Q^2)}.\label{eq:TLRKerrrNewman}
\end{eqnarray}
Using (\ref{eq:scalefactor}), the scale factor is 
\begin{equation}
	R=\frac{2 M (2 M^2 -Q^{2})}{J}.  
\end{equation}
This leads to the dimensionless left- and right-moving CFT temperatures
\begin{equation}
	\bar{T}_{L}=\frac{2 M^2-Q^{2}}{4 \pi  J},\quad \bar{T}_{R}=\frac{M \sqrt{M^2 -\left(\frac{J^{2}}{M^{2}}+Q^{2}\right)}}{2 \pi  J}.
\end{equation}
And the dimensionless central charges are determined to be $\bar{c}_{L,R}=12J$~\cite{ChenLiuJHEP2012}. Notably, even though $\bar{T}_{L}, S_{L}$ of dyonic Kerr-Sen BH reduce to Kerr-Newman BH when $P=0$, the right-moving sector $\bar{T}_{R}, S_{R}$ of dyonic Kerr-Sen BH do NOT reduce to Kerr-Newman results. The fundamental difference of CFT thermodynamics between dyonic Kerr-Sen BH and Kerr-Newman BH is essentially in the right-moving sector. Furthermore, as for dyonic Kerr-Sen BH, if one applies directly the left- and right-moving temperatures (\ref{eq:TLRKerrrNewman}) to its left- and right-entropies without taking into account the scale factor on the temperatures, the central charges obtained are given by 
	\begin{equation}
		c_{L,R} = 24 M(2M^2-Q^2)\label{eq:cLcRKerrNewmanwithdimension}
	\end{equation}
which also does not depend on the angular momentum explicitly.

\section{Static Case: dyonic dilaton BH}
\label{sec:static}
In this section, we consider the static case for dyonic Kerr-Sen BH that we call as dyonic dilaton BH. One can easily obtain this solution by setting $a=0$. The spacetime metric of this BH is given by
\begin{equation}
	ds^2 = - \frac{\Delta}{\varrho^2 } dt^2+  \frac{\varrho ^2}{\Delta}dr^2 + \varrho ^2d\theta^2 + \varrho ^2 \sin^2\theta d\phi^2,\label{eq:ddmetric} \
\end{equation}
where
\begin{equation}
	\Delta = r^2-2dr-2m(r-d) -k^2+p^2+q^2, 
\end{equation}
\begin{equation}
	\varrho^2 = r^2-2dr -k^2.\label{eq:metricfunctiondd}\
\end{equation}
When the electromagnetic charges vanish, we obtain Schwarzschild BH solution. Remarkably, even for a generic charged dyonic BH, we cannot have an extremal solution since $\varrho^2$ has a zero at $r_c = M+d = P^2/M$ for $M^2 =\frac{P^2+Q^2}{2}$, signifying a curvature singularity. This is a spherical singularity~(see \cite{Senjaya:2026bmd}) locating at exactly the same position as the extremal horizon, i.e., the root of $\Delta=0$, and thus there is no BH. 

For a non-extremal BH, the entropy is then given by
\begin{equation}
	S_\pm = \pi(r_\pm^2 - 2dr_\pm -k^2),\label{eq:entropydd}
\end{equation}
and the temperature is
\begin{equation}
	T_\pm = \frac{r_\pm-(m+d)}{2\pi(r_\pm^2 - 2dr_\pm -k^2)}.\label{eq:temperaturedd}
\end{equation}
The electric and magnetic potentials are still in the similar form as (\ref{eq:ElPot})-(\ref{eq:MagPot}). The first law of thermodynamics of dyonic dilaton BH is given by
\begin{equation}
	dM = T_\pm S_\pm  + \Phi_\pm dQ +\Psi_\pm dP,\label{eq:firstlawdd}
\end{equation}
and the Smarr formula now is
\begin{equation}
	M =2 T_\pm S_\pm + \Phi_\pm Q +\Psi_\pm P,\label{eq:Smarrdd}
\end{equation}
Moreover, the Christodoulou-Ruffini square mass formula in this case explicitly reads
\begin{equation}
	M^2 = \frac{A_\pm}{16\pi}+\frac{P^2+Q^2}{2}.\label{eq:Christodouloumassdd}
\end{equation}

Even for the non-extremal BH, since $r_{-}^2 - 2dr_{-} -k^2=0=\varrho^2(r_{-})$, the inner horizon coexists with the singularity and thermodynamic quantities associated with this horizon become unphysical. However, we can take an arbitrarily small $a\to 0$ limit and consider thermodynamics of both horizons. In such limit, the entropy product approaches zero while the temperature product becomes singular. The vanishing entropy product is actually caused by $S_-=S_L - S_R=0$ which denotes that the inner horizon entropy shrinks down to a spatial point or a surface of zero area. Moreover, the inner horizon temperature is divergent, implying that the equilibrium thermodynamic description completely breaks down at the inner horizon and we cannot treat $r_-$ as a stable physical thermal state due to the presence of singularity. This is different from the Reissner-Nordstr{\"o}m BH where the entropy product still depends on the electric charge and the temperature product is not singular \cite{ChenZhangJHEP2013}. The singularity of the temperature product corresponds to the singular dimensionless left- and right-moving CFT temperatures since the scale factor $R$ is singular. However, we can observe that the left- and right-moving CFT temperatures and entropies of the dyonic dilaton BH coincide to each other and non-vanishing as,
\begin{equation}
	T_{L,R}= \frac{1}{8\pi M},~~~	S_{L,R}=\pi(2M^2-P^2-Q^2),\label{eq:TLRSLRdd}
\end{equation}
Assuming Cardy entropy formula must be valid using above $T_{L,R}$ for this BH, one can derive the central charge
\begin{equation}
	c_{L,R}=24M(2M^2-P^2-Q^2).\label{eq:cLRdd}
\end{equation}
It is worth noting that the central charges of the dyonic Kerr-Sen BH (\ref{eq:cLcRwithdimension}) is similar to the static case above which do not depend on $J$ explicitly while its dimensionless central charges (\ref{eq:cLcR}) depend on $J$. It is because the temperatures employed to compute the dimensionless central charges (\ref{eq:cLcR}) for the rotating version are the dimensionless temperatures which have been re-scaled by the scale factor (\ref{eq:scalefactor1}), or generally (\ref{eq:scalefactor}). Since the scale factor is proportional to $1/(\Omega_+ - \Omega_-)$, it must depend on the angular momentum.
	
For static case, we directly employ the temperatures without the scale factor to find the central charges. This implies that the central charges for the static case (\ref{eq:cLRdd}) are not the dimensionless central charges, unlike the central charge for the rotating case (\ref{eq:cLcR}). We cannot employ the scale factor to the computation in the static case since it will cause the irregularity in the CFT temperatures resulting in vanishing central charges which are trivial. This is fair since our result can reduce to the results of the Schwarzschild BH case given in Refs. \cite{LoweSkanataJPA2012,ShajieeIJTP2016} when the charges are zero. The computation using hidden conformal symmetry in Refs. \cite{LoweSkanataJPA2012,ShajieeIJTP2016} shows that the Schwarzschild BH can be represented by a single CFT sector with the central charge $c_{Sch}=96M^3$ which is not the dimensionless one where the temperature applied to Cardy entropy is $T_{Sch}=T_H=1/8\pi M$. For this Schwarzschild BH case, the central charge can be interpreted as the total central charge from left and right sectors.

If one is interested to compute the dimensionless central charge, we refer to \cite{RopotenkoMPLA2016} where it is required to assume the existence of the intrinsic angular momentum. For instance, Schwarzschild BH possesses the instrinsic angular momentum as $J_{in}=2M^2$ which results in the dimensionless central charge $\bar{c}_{Sch}=12J_{in}=24M^2$. Furthermore, this intrinsic angular momentum is eventually proportional to the area of the BHs in Ref. \cite{RopotenkoMPLA2016} where $J_{in}=\frac{A_+}{8\pi}$. This relation that the dimensionless central charge is proportional to the outer horizon area of the BH is exhibited also in Ref. \cite{MeiJHEP2010} using the Kerr/CFT correspondence where 
\begin{equation}
\bar{c}_L = \frac{3A_+}{2\pi}. \label{eq:cLtoA}
\end{equation}	
Note that this is the result for the extremal BHs where the right-moving sector vanishes. When the right-moving sector is present, both sectors must have the equal central charges \cite{CastroPRD2010,SaktiGREG2019,SaktiNucPhysB2020,SaktiPhysDarkU2022}. The result in Ref. \cite{RopotenkoMPLA2016} is supported by the computation using soft hair method in Refs. \cite{AverinJHEP2019,AverinPRD2020} which generalizes the computation of the central charge that can be represented by two sectors in CFT. So, based on the relation (\ref{eq:cLtoA}), we argue that since the dimensionless central charge is actually proportional to outer horizon area of the BH, and the dimensionless left- and right- moving central charges are equal, the dimensionless central charges of the dyonic dilaton BH is given by 
\begin{equation}
\bar{c}_{L,R}=\frac{3A_+}{4\pi}=6(2M^2-P^2-Q^2).\label{eq:cLRddnew}
\end{equation}
Thus, even though the angular momentum vanishes in the static case, the dimensionless central charges remain non-zero and non-trivial.

\section{Hidden conformal symmetry and quasi-normal modes}
\label{sec:QNM}

\subsection{Hidden conformal symmetry}
Before studying the QNM spectrum on the dyonic Kerr-Sen BH background, we revisit the hidden conformal symmetry calculation on dyonic Kerr-Sen BH. It is worth noting that we use the spacetime metric (\ref{eq:dKSmetric}) instead of the metric in shifted radial coordinate as it has been computed in \cite{SaktiEPJC2023,SaktiPiyabutEPJC2024}. Now we consider massless neutral quantum scalar field where the solution has generic form
\begin{equation}
	\chi (t,r,\theta,\phi)=e^{-i\omega t+in\phi}P_l^n(\theta)R(r),
\end{equation}
where $\omega$ is the asymptotic energy and $n$ is the angular momentum of the scalar field. Plugging in above ansatz to the following Klein-Gordon equation,
\begin{equation}
	\nabla_{\alpha} \nabla^{\alpha}\chi = 0\label{eq:KG},
\end{equation}
one can find radial equation in low-frequency limit ($\omega M\ll 1$) that reads as
\begin{equation}
	\left[ \partial_r \left(\Delta \partial_r\right) + \frac{r_+ - r_-}{r - r_+} A + \frac{r_+ - r_-}{r - r_-} B -l(l+1) \right]  R(r) = 0 , \label{eq:radeq} \
\end{equation} 
where
\begin{eqnarray}
	&&A=\frac{\left[(r_+^2 -2dr_+ -k^2+a^2)\omega - an \right]^2}{(r_+ -r_-)^2},\nonumber\\
	&&B=-\frac{\left[(r_-^2 -2dr_- -k^2+a^2)\omega - an \right]^2}{(r_+ -r_-)^2}. \label{eq:AB}\
	\end{eqnarray}
Moreover, from $\Delta=0$, we can use $r_\pm^2 -2dr_\pm -k^2+a^2 =2m(r_\pm -d) -p^2 -q^2= 2mr_\pm-2p^2$ to write
\begin{eqnarray}
	&&A=\frac{\left[(2m r_+-2p^2)\omega - an \right]^2}{(r_+ -r_-)^2},\nonumber\\
	&&B=-\frac{\left[(2m r_- -2p^2)\omega - an \right]^2}{(r_+ -r_-)^2}, \label{eq:AB1}\
\end{eqnarray}
where the separation constant $l(l+1)$ is the eigenvalues on a sphere. One may expect that due to the electromagnetic duality, there is no ($-2p^2$)-term in radial wave equation. Nonetheless, the appearance of ($-2p^2$)-term in (\ref{eq:AB1}) naturally comes from the relation $\Delta=0$. However, ($-2p^2$)-term will vanish when we work on the shifted radial coordinate, $r\rightarrow r+ p^2/m$ or $r\rightarrow r+ d$ which is related to the gauge choice \cite{WuWuWuYuPRD2021}. The angular equation in low-frequency limit is just a Laplacian on the two-sphere \cite{CastroPRD2010}. The main reason we do the computation of the hidden conformal symmetry is to show that the left- and right-moving temperatures are explicitly the same with the results from the thermodynamic method as given in Eq. (\ref{eq:dimTLR}). Thus to reveal the hidden conformal symmetry, we perform the following coordinate transformations
\begin{eqnarray}
	&&\omega^+ = \sqrt{\frac{r-r_+}{r-r_-}}\mathrm{e}^{2\pi \bar{T}_R \phi + 2 \bar{n}_R t}, \label{eq:con1}\\
	&&\omega^- = \sqrt{\frac{r-r_+}{r-r_-}}\mathrm{e}^{2\pi \bar{T}_L \phi + 2 \bar{n}_L t}, \label{eq:con2}\\
	&&y = \sqrt{\frac{r_+ -r_-}{r-r_-}}\mathrm{e}^{\pi (\bar{T}_L +\bar{T}_R) \phi + (\bar{n}_L + \bar{n}_R) t}, \label{eq:con3}\
\end{eqnarray}
where $\bar{n}_{L,R}$ is the conserved charge in CFT. From those conformal coordinates, we can define three locally conformal operators in terms of the new conformal coordinates $\omega^+,\,\omega^-$ and $y$ as
\begin{eqnarray}
	&& H_1 = i \partial_+, \label{vec1}\\
	&& H_0 = i \left(\omega^{+}\partial_+ + \frac{1}{2}y\partial_y \right), \label{vec2}\\
	&& H_{-1} = i \left(\omega^{+2}\partial_+ + \omega^{+}y\partial_y - y^2 \partial_- \right), \label{vec3}
\end{eqnarray}
as well as 
\begin{eqnarray}
	&& \bar{H}_1 = i \partial_-, \label{vvec1}\\
	&& \bar{H}_0 = i\left(\omega^{-}\partial_- + \frac{1}{2}y\partial_y \right), \label{vvec2}\\
	&& \bar{H}_{-1} = i \left(\omega^{-2}\partial_- + \omega^{-}y\partial_y - y^2 \partial_+ \right). \label{vvec3}
\end{eqnarray}
Note that we have defined $\partial_\pm = \frac{\partial}{\partial \omega^\pm}$. Each set of operators given in Eqs. (\ref{vec1})-(\ref{vec3}) and (\ref{vvec1})-(\ref{vvec3}) satisfies the $ SL(2,R) $ Lie algebra,
\begin{eqnarray}
	\left[H_0,H_{\pm 1} \right] = \mp iH_{\pm 1}, ~~~ \left[H_{-1},H_1 \right]=-2iH_0.
\end{eqnarray}
Hence the quadratic Casimir operator can be formed from any of two sets of operators as
\begin{eqnarray}
	\mathcal{H}^2 &=& \bar{\mathcal{H}}^2 = - H_0^2 + \frac{1}{2}(H_1 H_{-1} + H_{-1} H_{1})\nonumber\\
	&=& \frac{1}{4}(y^2 \partial_y^2 - y\partial_y)+y^2\partial_+ \partial_-. \label{eq:quadraticCasimir}\
\end{eqnarray}
Now the radial wave equation (\ref{eq:radeq}) can be represented as the Casimir operator (\ref{eq:quadraticCasimir}) that represents $SL(2,R)\times SL(2,R)$ isometry. Simply, we can write
\begin{eqnarray}
	\mathcal{H}^2 R(r)=\bar{\mathcal{H}}^2 R(r)= l(l+1) R(r), \label{quadraticCasimirEq}
\end{eqnarray}
that is satisfied when the dimensionless left- and right-moving CFT temperatures are given by Eq. (\ref{eq:dimTLR}).

\subsection{QNM Spectra}
In order to calculate the QNM spectra from radial equation (\ref{eq:radeq}) in low-frequency limit, we can apply the near-horizon/far-region matching technique which has been applied to Kerr black holes \cite{CastroPRD2010,KehagiasPerronebJCAP2023,CveticRodriguezPRD2025,CveticPerryPRD2026} where the QNM spectra can be read off from the poles of the absorption cross-section. In the low-frequency limit, the radial wave equation maps to a hypergeometric differential equation, where the QNM boundary conditions are satisfied by selecting the poles of the scattering coefficients. Thus we can find the analytical form of the QNM spectra in this limit. We want to emphasize that this calculation is specifically applied only in the near-horizon region, yet we also consider the wave solution which satisfies the asymptotic infinity condition. For a full QNM spectrum (including high-frequency or "overtone" modes), one must typically resort to numerical methods, for instance, the continued fraction method \cite{LeaverPRD1990} or WKB approximation \cite{HuangLiPRD2026}, as the hidden conformal symmetry technically does not describe the full radial equation of the BH spacetime.

In order to obtain the solutions to the radial wave equation (\ref{eq:radeq}), we apply the coordinate transformation
\begin{equation}
z = \frac{r-r_+}{r-r_-},\label{eq:zcoord}
\end{equation} 
such that $r_+ \leq r \leq \infty$ corresponds to $0 \leq z \leq 1$. In terms of new coordinate $z$, the radial wave equation becomes
\begin{equation}
\left[z(1-z)\partial^2_z + (1-z)\partial_z +\frac{A}{z}-B-\frac{l(l+1)}{1-z}\right]R(r)=0.\label{eq:radialeqz}
\end{equation}
The radial wave equation (\ref{eq:radialeqz}) possesses two independent solutions with ingoing and outgoing boundary conditions. Since we focus on the near-horizon region, we can find the QNM spectra with the ingoing boundary condition at $z=0  ~(r=r_+)$ \cite{SakalliMPLA2013,SakalliMirekhtiaryASS2014,SakalliPRD2016}. The ingoing solution is given by
\begin{equation}
	R(r)=c_1 \left(\frac{r-r_+}{r-r-}\right)^{c_2}(r-r_-)^{-l-1} {F}\left(c_3,c_4;c_5;z\right),\label{eq:radialsol}
\end{equation}
where ${F}\left(c_3,c_4;c_5;z\right)$ is a hypergeometric function, $c_1$ is just an integration constant, and
\begin{equation}
	c_2 = -i\frac{\omega -n\Omega_+}{2\pi T_+}, ~~c_3=1+l-\frac{i\omega}{4\pi T_R}+\frac{in\Omega_+}{2\pi T_+},\nonumber\
\end{equation}
\begin{equation}
	c_4 =1+l -\frac{i\omega}{4\pi T_L}, ~~c_5=1-\frac{i\omega}{2\pi T_+}+\frac{in\Omega_+}{2\pi T_+}.\nonumber\
\end{equation}
The outgoing solution at $z=1$ to Eq. (\ref{eq:radialeqz}) is given by
\begin{equation}
	R_{out}(r)=c_1 \left(\frac{r-r_+}{r-r-}\right)^{-c_2}(r-r_-)^{-l-1} {F}\left(c_3,c_4;c_5;z\right),\label{eq:radialsoloutgoing}
\end{equation}
in which we will not use it in the computation of the absorption cross-section.

To calculate the absorption cross-section for a wave coming from infinity towards the event horizon $(r=r_+)$, we need to apply the ingoing solution (\ref{eq:radialsol}). At large $r$ or $z\to1$, we can employ the transformation law for hypergeometric functions \cite{CveticRodriguezPRD2025,CveticPerryPRD2026,KehagiasPerronebJCAP2023},
\begin{eqnarray}
F(a,b;c;z)&=&\frac{\Gamma(c)\Gamma(c-a-b)}{\Gamma(c-a)\Gamma(c-b)}\nonumber\\
&\times& F(a,b;a+b-c+1;1-z)\nonumber\\
 &+&(1-z)^{c-a-b}\frac{\Gamma(c)\Gamma(a+b-c)}{\Gamma(a)\Gamma(b)}\nonumber\\
 &\times&F(c-a,c-b;c-a-b+1;1-z).\nonumber\\
 &~&  \label{eq:hypergeometrytrans}\
\end{eqnarray}
Note that $F(a,b;c;0)=1$ and $1-z=\frac{r_+ -r_-}{r- r_-}\rightarrow \frac{r_+ -r_-}{r}$. As $r\rightarrow \infty$ or $r\gg M$, the first term in Eq. (\ref{eq:hypergeometrytrans}) behaves as $r^{-1-l}$ while the second term behaves as $r^l$ or we can write $R(r)\sim Er^{-1-l}+ Dr^l $. Thus the first term can be neglected compared to the second term. Now the ingoing solution satisfies the asymptotic region and the near-horizon condition. From Eq. (\ref{eq:hypergeometrytrans}) by setting $a\rightarrow c_3, b\rightarrow c_4, c\rightarrow c_5$, we can  find the coefficient $D$ on $R(r)\sim Dr^l$ which reads as
\begin{equation}
	D= \frac{\Gamma(1+2l)\Gamma\left(1-i\frac{\omega -n\Omega_+}{2\pi T_+}\right)}{\Gamma\left(1+l -\frac{i\omega}{4\pi T_L}\right)\Gamma\left(1+l-\frac{i\omega}{4\pi T_R}+\frac{in\Omega_+}{2\pi T_+}\right)}.\label{eq:D}
\end{equation}
The above constant encodes the information of the absorption cross-section which is given by \cite{SaktiEPJC2023,SaktiPiyabutEPJC2024,CveticRodriguezPRD2025,CveticPerryPRD2026}
\begin{eqnarray}
	P_{abs} &\sim& |D|^{-2} =\sinh\left(\frac{\omega -n\Omega_+}{2T_+}\right)\bigg|\Gamma\left(1+l -\frac{i\omega}{4\pi T_L}\right)\bigg|^2\nonumber\\
	&\times& \bigg|\Gamma\left(1+l-\frac{i\omega}{4\pi T_R}+\frac{in\Omega_+}{2\pi T_+}\right)\bigg|^2.\label{eq:Pabs}\
\end{eqnarray}
The QNM spectra can be determined from the poles of the absorption cross-section. The poles of gamma functions are given when the value in the bracket is zero or negative integers. Thus from the absorption cross-section (\ref{eq:Pabs}), we can find two different QNM spectra as follows
\begin{eqnarray}
	\omega_L &=& -4\pi i T_L (1+l +n_L), \label{eq:QNML}\\
	\omega_R &=& -4\pi i T_R (1+l+n_R) + \frac{2n\Omega_+ T_R}{T_+}, \label{eq:QNMR}\
\end{eqnarray}
where $n_{L,R}=0,1,2,...$. As we have mentioned in the beginning of this section, we emphasize that the QNM spectra given by (\ref{eq:QNML}) and (\ref{eq:QNMR}) are valid in the low-frequency limit and near-horizon regime  where the duality between BH in the near-horizon region and 2D CFT holds. In addition, the family of modes given by the second family of modes (\ref{eq:QNMR}) has identical frequencies to those of \cite{HodPLB2008} where the QNM spectra take the form
\begin{equation}
	\omega = -2\pi i T_R (1+l+n) + 2n\Omega_+.\label{eq:QNMresonance}
\end{equation}
which is valid in the regime of Im$(\omega)\ll$ Re$(\omega) \ll 1/M$ Note that this does not apply to the first family of modes (\ref{eq:QNML}). 

It is plausible that the first family of modes (\ref{eq:QNML}) is purely imaginary QNM spectrum as it was explained firstly by Maggiore \cite{MaggiorePRL2008}. To explain this, Maggiore \cite{MaggiorePRL2008} described a QNM spectrum as a damped harmonic oscillator $f(t)$ satisfying the following equation,
\begin{equation}
\ddot{f} +2\gamma_0 \dot{f}+\omega_0^2 f = F(t),
\end{equation}
where $\gamma_0, \omega_0, F(t)$ are the damping constant, proper frequency of the oscillator, and external force, respectively. With the boundary condition $f(t)=0$ for $t<0$ and the external force is just a delta function $F(t)=\delta(t)$, the solution is given by
\begin{equation}
f(t)=\frac{i}{\omega_+ -\omega_-}\left(e^{-i\omega_+ t}-e^{-i\omega_- t}\right),
\end{equation}
where $\omega_\pm = \pm \sqrt{\omega_0^2 -\gamma_0^2}-i\gamma_0\equiv\omega_{R}+i\omega_{I}$. Since $\omega_{R} =\sqrt{\omega_0^2 - \omega_{I}^2}$, so $\omega_{0}\simeq \omega_{R}$ is only true for the long-lived modes with $\gamma_0\ll \omega_0$. For most BH QNM spectrum, the highly excited modes satisfy $\omega_{I}=\gamma_0\gg \omega_{R}$, and consequently $\omega_0 \simeq \omega_{I}$. This will become important in the entropy quantization of the BH in the next Section. 

\section{Bekenstein-Hod conjecture}
\label{sec:BekensteinHod}

The study of Kerr/CFT correspondence at first was applied only in the extremal Kerr BH where the duality can be shown from the entropy matching between Cardy entropy formula from 2D CFT and Bekenstein-Hawking entropy \cite{GuicaPRD2009,SaktiSurosoIJMPD2018,SaktiSurosoEPJPlus2019,SaktiAnnPhys2020,SaktiPhysDarkU2021}. Furthermore, to support in the duality between 2D CFT and BH, Ref. \cite{CastroPRD2010,SaktiNucPhysB2020,SaktiGREG2019,SaktiPhysDarkU2022} studied the matching of absorption cross-section for generic non-extremal black hole in the low-frequency limit. Moreover, since the poles of the absorption cross-section encodes the approximated QNM spectrum in the low-frequency limit, Ref. \cite{CveticRodriguezPRD2025,CveticPerryPRD2026} also computed the QNM spectrum for Kerr BH from supergravity theory which corresponds to the Bekenstein-Hod conjecture. However, these work cannot be generalized to the exact high overtone number of QNM spectra.

We want to see the relation between the quantization of the entropy and the QNM spectra which is based on the Bekenstein-Hod conjecture. The Bekenstein-Hod conjecture is a theoretical proposal that links the classical "ringing" of a BH, i.e. QNM spectrum to the quantum nature of BH area. Bekenstein proposed that the BH area is an adiabatic invariant and should be quantized \cite{Bekenstein1972,Bekenstein1973,Bekenstein1974}. Separately, Hod suggested that the highly damped QNM spectrum of a BH represents the fundamental frequencies of the BH horizon \cite{HodPRL1998}. Drawing on the Bohr correspondence principle, Hod argued that these classical frequencies should dictate the spacing of the quantum energy (and therefore area) levels of BH.

Now imagine that we perturb the BH by dropping in a single quantum of energy $\delta M$ while keeping other parameters fixed. This quantum perturbation can come from either branch of the QNM spectra, the left-moving or the right-moving family. For the right-moving case, we first take $n=0$, so the energy delivered to the horizon is then just $|\omega_R|$ where $dE_R=dM=|\omega_R|$. The same logic applies for the left-moving branch with $|\omega_L|$ playing the analogous role. Plugging this into the first law written for the right (or left) sector shows that $\delta S_R$ (or $\delta S_L$) always comes out as $2\pi \times$ some integer, never a fractional multiple. For example, from the right-moving of modes (\ref{eq:QNMR}) we set $l=0,n_R=0$, so we have $|\omega_R|=4\pi T_R$. Since other parameters are held fixed, except $dM=dE_R$, we have $dS_R =dE_R/T_R = 4\pi$. Thus, since the total horizon entropy splits as $S_+=S_R+S_L$, this quantization on each sector immediately carries over to the sum, so the entropy (or area) of the BH must be also quantized.

\section{Summary}
\label{sec:summary}
In this work, we have investigated the inner and outer horizon thermodynamics of the dyonic Kerr-Sen black hole in four-dimensional Einstein-Maxwell-Dilaton-Axion theory, which carries mass $M$, angular momentum $J$, electric charge $Q$, and magnetic charge $P$. We have carried out the thermodynamic quantities such as entropy, temperature, angular velocity, electric potential, and magnetic potential on both the event horizon and the Cauchy horizon. We have explicitly verified the first law and Smarr formula on each horizon.

We have computed the entropy product and sum for the dyonic Kerr-Sen black hole and demonstrated that the entropy product $S_-S_+ = (2\pi J)^2$ is mass-independent and depends solely on the angular momentum, establishing the universality of the entropy (or area) product. This universality is equivalent to the condition $T_+ S_+ + T_- S_-=0$, which we also verified. From the entropy sum and product, we derived Penrose-like entropy bounds for both horizons.

Exploiting the universal entropy product, we constructed the dual two-dimensional CFT thermodynamics through the Kerr/CFT correspondence via the thermodynamic method. The dimensionless central charges of the left- and right-moving sectors were found to be $c_L=c_R=12J$, identical to those of the Kerr, Kerr-Newman, and Kerr-Sen black holes. In the extremal limit, the microscopic Cardy entropy $S_{CFT}=2\pi J$ is in perfect agreement with the macroscopic Bekenstein-Hawking entropy. Comparing to the results from Kerr-Newman BH, when we turned off the magnetic charge of dyonic Kerr-Sen BH, the left-moving sector has similar results while the right-moving sector is fundamentally different.

We also carried out the static solution from the dyonic Kerr-Sen BH and its thermodynamics. It has been shown that the entropy product vanishes due to the fact that $S_-=0$, implying that the inner horizon area is shrinking to zero. The limit $a\to 0$ brings singularity to coexist with the inner horizon and the dyonic Kerr-Sen BH has only one horizon in the static limit. Thus one could not reproduce the regular dimensionless CFT temperatures and non-vanishing dimensionless central charges from that of the rotating case. However, from the thermodynamic method, we have found that the left- and right-moving CFT temperatures are regular. Furthermore, by using Cardy entropy formula, it has been found that $c_{L,R}=24M(2M^2-P^2-Q^2)$ which is similar to those of its rotating version which reflects the non-vanishing left- and right-sectors in 2D CFT. Yet, these central charges are not the dimensionless ones but it still reduces to the result for Schwarzschild BH \cite{LoweSkanataJPA2012,ShajieeIJTP2016}. Moreover, due to the fact that the dimensionless central charges is proportional to the outer horizon area of the BH, we argued that the dimensionless central charges of the dyonic dilaton BH are given by $\bar{c}_{L,R}=6(2M^2-P^2-Q^2)$ in which both sectors must be equal. Thus, although the angular momentum is zero in the static case, the dimensionless central charges remain non-zero and non-trivial.

Finally, we revisited the hidden conformal $SL(2,R)\times SL(2,R)$ symmetry of the radial scalar wave equation in the low-frequency limit, confirming that the CFT temperatures derived thermodynamically are consistent with those from the hidden conformal symmetry analysis. From the poles of the greybody absorption cross-section, we obtained two families of QNM spectra providing a complete analytic expression for the QNM spectra of the dyonic Kerr-Sen black hole in terms of its physical parameters. Based on the Bekenstein-Hod conjecture, perturbing a BH with a single quantum of energy, whether it comes from the left or right QNM family, always changes the corresponding entropy by some integer multiple by $2\pi$. Since the total horizon entropy is just the sum of the two sectors of entropy, this corresponds to the quantization of BH entropy or area.

\begin{acknowledgments}
	This research is supported by the Second Century Fund (C2F) and C2F research abroad scholarship, Chulalongkorn University, Thailand. \\
	
	{\bf Data Availability Statement:} ~Data sharing not applicable to this article as no datasets were generated or analysed during the current study.
\end{acknowledgments}



\end{document}